\documentclass[runningheads]{llncs}
\usepackage{xcolor}
\usepackage{fancyvrb}

\usepackage{tcolorbox}
\tcbuselibrary{listings, breakable} 

\usepackage{inconsolata} 

\tcbuselibrary{listings}
\tcbuselibrary{breakable} 

\definecolor{codebg}{RGB}{248, 249, 250}       
\definecolor{codeborder}{RGB}{222, 226, 230}   
\definecolor{keycolor}{RGB}{215, 58, 73}       
\definecolor{strcolor}{RGB}{3, 47, 98}         
\definecolor{commentcolor}{RGB}{106, 115, 125} 

\newtcblisting{promptbox}[1][]{
    enhanced,
    colback=codebg,
    colframe=codeborder,
    boxrule=0.5pt,
    arc=3pt, 
    top=4pt, bottom=4pt, left=6pt, right=6pt,
    title=#1,
    coltitle=black,
    fonttitle=\sffamily\bfseries\small,
    attach boxed title to top left={yshift=-1.5mm, xshift=4mm},
    boxed title style={colback=white, colframe=codeborder, boxrule=0.5pt, arc=2pt},
    listing only,
    listing options={
        language=Python,
        basicstyle=\ttfamily\small, 
        keywordstyle=\color{keycolor}\bfseries,
        stringstyle=\color{strcolor},
        commentstyle=\color{commentcolor}\itshape,
        showstringspaces=false,
        breaklines=true
    }
}
\newif\ifshowcomments
\showcommentsfalse 

\ifshowcomments
    \newcommand{\TA}[1]{\textcolor{blue}{\textbf{*TA*}: #1}}
    \newcommand{\SC}[1]{\textcolor{brown}{\textbf{*SC*}: #1 }}
    \newcommand{\NN}[1]{\textcolor{magenta}{\textbf{*NN*}: #1 }}
    \newcommand{\AH}[1]{\textcolor{orange}{\textbf{*AH*}: #1 }}
    \newcommand{\todo}[1]{\textcolor{red}{\textbf{(TODO)}: #1}}
\else
    \newcommand{\TA}[1]{}
    \newcommand{\SC}[1]{}
    \newcommand{\NN}[1]{}
    \newcommand{\todo}[1]{}
    \newcommand{\AH}[1]{}
\fi

\usepackage{xspace}

\newcommand{\schema}{\texttt{schema.org}\xspace}
\newcommand{\structured}{Semantic Agent\xspace}
\newcommand{\baseline}{Baseline Agent\xspace}

\usepackage{booktabs}
\usepackage{amsmath}  

\usepackage[T1]{fontenc}
\usepackage{graphicx}
\begin{document}
\title{Do Data Agents Need Semantic Metadata? A Comparative Study in Agentic Data Retrieval}
\titlerunning{Do Data Agents Need Semantic Metadata?}
%
\author{Shiyu Chen
\and
Tarfah Alrashed
\and
Alon Halevy
\and
Natasha Noy}

%
\authorrunning{S. Chen et al.}
%
\institute{Google, Mountain View, CA, USA \\
\email{\{shiyuc,tarfah,halevy,noy\}@google.com}}
\maketitle              

\begin{abstract}
In the era of autonomous agents, machine-actionable data is critical for data-driven  workflows. For more than a decade, semantic metadata like \schema has anchored the FAIR principles (Findable, Accessible, Interoperable, and Reusable) for machine-actionable data and enabled discovery tools like Google Dataset Search. However, the rise of Large Language Models (LLMs) capable of navigating the unstructured web raises a fundamental question: Is semantic metadata still necessary for agentic data discovery, or can agents reliably retrieve actionable data directly from the  web?
We present a comparative analysis of agentic data retrieval across two distinct environments: a \baseline searching billions of open-web documents, and a \structured leveraging a corpus of 90 million datasets using \schema. We deploy an ``LLM-as-a-judge'' evaluation pipeline, mapped directly to the FAIR principles, to assess the semantic relevance, data accessibility, and computational utility of the retrieved data.
Our results reveal a clear divergence. The \structured excels at retrieving actionable data, achieving a 44.9\% higher precision for metadata-rich registries and a 46.6\% higher precision for pages with machine-readable downloads among its returned results. Conversely, the \baseline frequently suffers ``Last-Mile Utility'' failures, retrieving prose-heavy pages (20.1\% of results) and portal landing pages (8.5\%) rather than actual data pages. While the \baseline achieves higher coverage by answering 40\% more questions, the \structured delivers greater accuracy, achieving 65.7\% higher overall precision in retrieving FAIR-compliant datasets. We conclude that while unstructured retrieval supports broad exploratory tasks, structured ecosystems remain the indispensable foundation for reliable, execution-oriented autonomous workflows.
\end{abstract}

\keywords{Semantic Metadata  \and Data Discovery  \and Large Language Models  \and Autonomous Agents \and Retrieval-Augmented Generation }

\section{Introduction}
\label{sec:introduction}

As the world becomes increasingly data-driven, researchers and practitioners rely heavily on open data to answer scientific questions and to understand complex phenomena~\cite{zuiderwijk2020drives}. This deep reliance on data has catalyzed a shift in scientific communication, making publication of primary data the norm across many scientific disciplines~\cite{tedersoo2021data}. Users do not merely need to read about data; they frequently need to find it and to act upon it---for example, to conduct independent analyses of primary sources or to replicate experimental results.

Today, there are tens of millions of datasets scattered across the Web. Finding specific, usable data in this vast, unstructured space is a massive challenge~\cite{dataset_or_not}. We built Google Dataset Search~\cite{dataset_search} specifically for this use case. Recognizing that traditional Web search engines are optimized for narrative prose rather than structured data, we relied on semantic metadata---specifically \schema~\cite{schema_guha} and W3C DCAT~\cite{w3c_dcat3}\cite{albertoni2024dcat}---to construct a dedicated vertical search engine. By incentivizing publishers to add explicit semantic metadata for the datasets, this structured ecosystem organized the chaotic subset of the Web describing datasets into a reliable, deterministic corpus for data discovery~\cite{dataset_search_by_the_numbers}.

Historically, semantic metadata facilitated the FAIR principles—ensuring data is Findable, Accessible, Interoperable, and Reusable~\cite{wilkinson2016fair}. These principles emphasize machine-actionability: the capacity for computational systems to utilize this data with minimal human intervention. Today, the emergence of autonomous agents~\cite{yao2023react,10.1145/3586183.3606763,NEURIPS2023_d842425e} transforms FAIR from a human-centric best practice into a technical requirement. For an agent, \textit{findability} is merely the starting point. Bridging the gap to data execution requires machine-actionable \textit{accessibility}—such as direct APIs and machine-readable formats—so agents can process the data seamlessly. Furthermore, \textit{interoperability} via semantic attributes ensures agents understand dataset schemas, streamlining the integration of extracted data into downstream workflows. Finally, \textit{reusability} provides the provenance and licensing context needed to verify whether or not it is appropriate for the agent to use the asset, preventing the blind extraction of unverified data. Ultimately, the advent of agents does not diminish the need for FAIR data; rather, it makes FAIR the essential foundation for reliable autonomous workflows.

Simultaneously, modern LLMs have demonstrated a remarkable capability to navigate the unstructured web directly. A dataset-discovery agent built entirely on traditional web search~\cite{nakano2021webgpt,ICLR2024_4410c071} can extract context straight from HTML prose, effectively bypassing traditional metadata structures. 
This mechanism  stands in  contrast to the structured ecosystem of the Semantic Web, where semantic metadata constitutes the ground truth---unambiguous, machine-readable statements guaranteeing schema and provenance. The capability with which LLMs navigate the unstructured pages raises a critical question: if an agent locates data without these semantic guarantees, is that asset truly FAIR and actionable, or simply findable?

This friction drives our core research question: In an agent-led system, do we still need semantic metadata to empower data discovery, or can LLMs natively bridge the gap through unstructured retrieval?

To answer this question, we present a comparative analysis of agentic data retrieval across two environments: Using a similar agent architecture, we evaluate a \textbf{\baseline} (billions of open-web documents accessed via general web search) against a \textbf{\structured}. The \structured  operates over a corpus of 90 million dataset metadata records from Google Dataset Search, which relies on the presence of \schema semantic markup (Section~\ref{sec:system_architecture}).

To assess retrieval quality in a scalable way, we deploy a multi-dimensional ``LLM-as-a-judge'' pipeline~\cite{zheng2023judging} mapped directly to the FAIR principles. Using keyword-based queries from the NTCIR-16 Data Search benchmark~\cite{tao2022overview}, our autoraters evaluate the retrieved assets across semantic relevance, data accessibility, and computational utility (Section~\ref{sec:evaluation_method}).

The following \textbf{research questions} drive our comparative framework:

\begin{itemize}
    \item \textbf{Data Actionability}: Does semantic metadata provide an advantage in surfacing actionable datasets compared to unstructured web search systems? \textit{Hypothesis:} \structured will yield a significantly higher success rate for locating actionable, downloadable datasets.

    \item \textbf{The ``Last Mile'' Utility}: When unstructured web search fails in identifying actionable data, what is the reason? Is it contextual divergence---where the model loses focus on the data objective in favor of the surrounding prose---or explicit hallucination? \textit{Hypothesis:} \baseline often fails the ``last mile'' of data retrieval, landing on web pages requiring further navigation or complex extraction rather than immediate data payloads.

    \item \textbf{Exploratory Breadth vs. Precision}: How do the two agents balance exploratory recall against retrieval precision? \textit{Hypothesis:} The \baseline will achieve higher overall query recall by acting as a broader discovery superset for niche topics lacking semantic markup, whereas the \structured will deliver significantly higher precision and more focused results.
\end{itemize}

In summary, this paper makes the following contributions:

\begin{itemize}
    \item 
    We provide a comparative study evaluating the end-to-end utility of a semantic metadata ecosystem for agentic data retrieval.
    \item 
    We introduce a multi-dimensional, LLM-as-a-judge evaluation pipeline that translates the human-centric FAIR data principles into metrics for autonomous systems, extending evaluation beyond traditional semantic relevance. 
    \item 
    We publish the exact prompts used by our autoraters to ensure full transparency and foster reproducibility of our LLM-as-a-judge methodologies.
    
\end{itemize}

\section{Related Work}


The challenge of agentic data retrieval sits at the intersection of RAG, dataset discovery, and Semantic Web technologies. Despite the reasoning power of LLMs, integrating them with external search remains a research challenge, specifically regarding data reliability and utility.

\subsubsection{Autonomous Agents and Web Retrieval}

Autonomous agents built on RAG~\cite{lewis2020retrieval} leverage frameworks like ReAct~\cite{yao2023react}, Toolformer~\cite{NEURIPS2023_d842425e}, and ToolLLM~\cite{qin2023toolllm} for multi-step reasoning and tool execution. For web tasks, systems such as WebGPT~\cite{nakano2021webgpt} and WebArena~\cite{ICLR2024_4410c071} demonstrate that agents can navigate raw HTML to fulfill complex intents, but DOM traversal introduces high computational overhead and boilerplate noise that degrades reasoning~\cite{gur2023understanding}. Although distilling pages into Markdown or accessibility trees~\cite{ICLR2024_4410c071,wang2024browsergym} and pruning HTML~\cite{zhuang2024htmlrag} improve context management, agents with multi-step navigation still suffer from mis-grounding and poor state tracking~\cite{aghzal2026agent}. To bypass these bottlenecks, research has pivoted toward explicit API-calling for hallucination-free execution~\cite{patil2023gorilla}. Building on these findings, rather than optimizing multi-step scraping for data discovery, we restrict our agents to the search tools. By comparing agents querying different web ecosystem endpoints, we demonstrate that the metadata endpoint natively solves ``last mile'' failure of data retrieval. This approach sidesteps the computational overhead of multi-page scraping entirely while landing on a page with actionable data.




\subsubsection{Evaluating Dataset-Discovery Engines}
Unlike general web search, dataset retrieval requires meeting specific spatial, temporal, and format constraints~\cite{dataset_search,chapman2020dataset}. The NTCIR tasks~\cite{tao2022overview} formalized ad-hoc retrieval evaluation, yet focused primarily on matching natural language queries to metadata for human consumption. As consumers shift from humans to autonomous agents, \textit{findability} becomes a dead end without machine-actionable \textit{accessibility}. Our work extends the FAIR principles~\cite{wilkinson2016fair} into the agentic domain, evaluating the Google Dataset Search ecosystem~\cite{dataset_search,dataset_or_not,Sostek2024Discovering} not merely as a discovery tool, but as a critical infrastructure layer that solves the operational bottlenecks of autonomous execution.

\subsubsection{Automated Evaluation and LLM-as-a-Judge}
Agentic tasks are inherently open-ended. Thus, traditional exact-match metrics and rigid lexical comparisons fall short of capturing true task success~\cite{liu2023agentbench}. Consequently, the field has increasingly adopted the ``LLM-as-a-judge'' paradigm~\cite{zheng2023judging}, leveraging the zero-shot reasoning capabilities of strong foundational models to evaluate complex outputs against specialized rubrics. Frameworks such as MT-Bench~\cite{zheng2023judging} and G-Eval~\cite{liu2023geval} have demonstrated that LLM evaluators achieve high correlation with human annotators, particularly for nuanced semantic tasks. In the context of data retrieval, human evaluation of FAIR compliance is prohibitively slow and subjective. By explicitly mapping our multi-dimensional LLM-as-a-judge pipeline to the FAIR principles, our methodology extends the automated evaluation paradigm beyond conversational fluency into the rigorous assessment of operational data utility.

\subsubsection{Structured Metadata and the Semantic Web in LLMs.}
While pairing LLMs with Knowledge Graphs (KGs) reduces hallucinations~\cite{pan2024unifying}, existing evaluations often rely on closed, domain-specific KGs~\cite{jiang2023structgpt,edge2025from} or static RAG frameworks where non-textual webpage metadata is injected primarily as a pre-filtered prompt feature~\cite{chiang2024metadata}. These designs reduce the LLM to a static synthesizer of provided evidence rather than an autonomous agent tasked with active information discovery.
Our work evaluates agent behavior across two architectures. In the \baseline, the agent actively navigates noisy search results to uncover real datasets. In the \structured, \schema metadata acts as a high-precision corpus filter for datasets, streamlining the high-entropy task of autonomous discovery. Comparing these modes isolates how semantic standards mitigate contextual divergence during autonomous data retrieval.

\section{System Architecture \& Experimental Setup}
\label{sec:system_architecture}



To isolate the impact of semantic metadata on dataset discovery, we designed two nearly identical workflows. Rather than analyzing granular variables like index size, we treat \schema metadata as a filter that fundamentally shapes corpus composition, search scale, and tool-execution logic.

\begin{figure}[h!]
    \centering
    \caption{Comparative System Architecture. Similar agent logic is evaluated across unstructured \baseline and \structured dataset search environments. Both feed a unified, FAIR-aligned evaluation of relevance, accessibility, and utility.}
    \includegraphics[width=\linewidth]{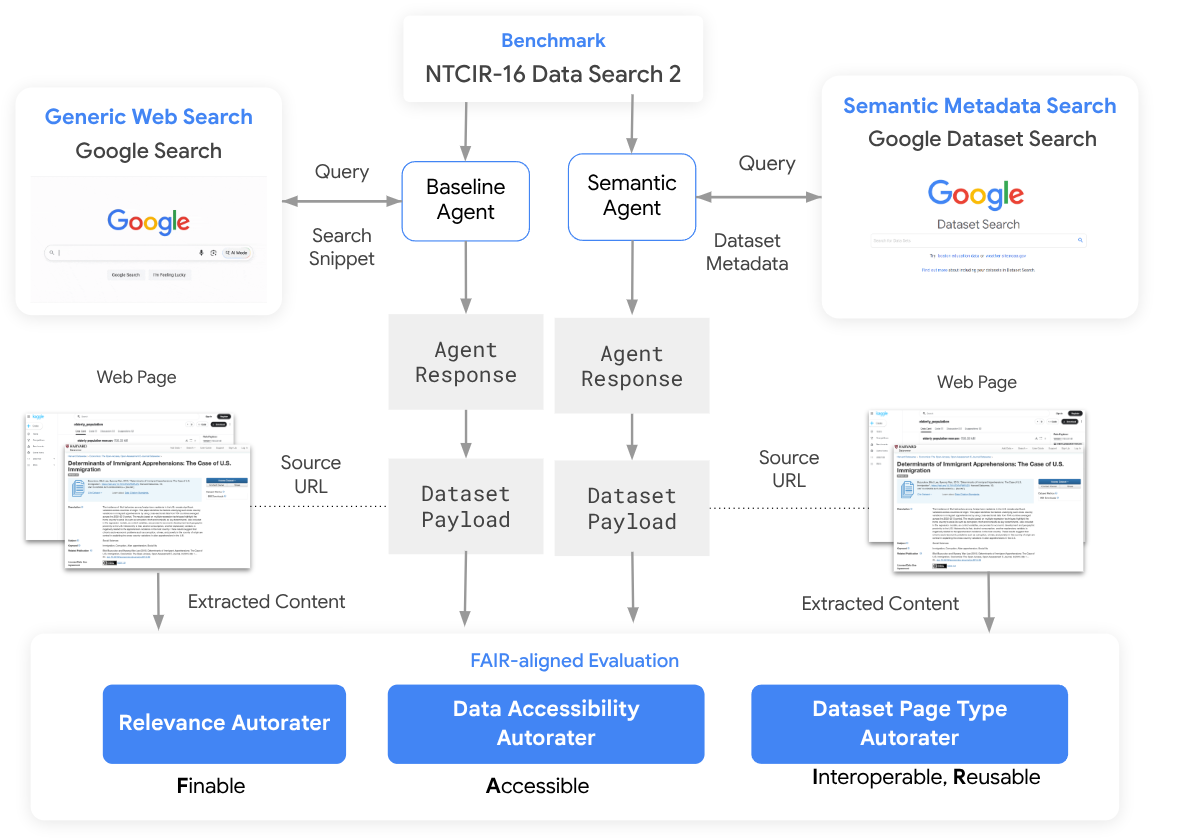}
    \label{fig:system-architecture}
\end{figure}

\subsection{Agentic Framework and Setup}

To ensure experimental parity, we contrast a \structured against a \baseline using identical underlying architectures (Figure~\ref{fig:system-architecture})---specifically, the Agent Development Kit (ADK)~\cite{google_adk} powered by Gemini 2.5 Pro. We isolate the target corpus and its respective search ecosystem as the primary independent variables. The agent prompts are highly symmetrical and tailored solely to optimize retrieval for their respective search environments, reflecting how users naturally interact with these platforms \footnote{prompts: \url{https://doi.org/10.6084/m9.figshare.32141311}}.

\begin{itemize}
    \item \textbf{Baseline Agent (Unstructured Corpus):} The agent queries the Google Search index via a standard search tool. Representing the unstructured web ecosystem, this baseline corpus is a superset of tens of billions of web documents, where inclusion relies purely on general crawlability. To optimize data retrieval, the agent explicitly appends data-seeking keywords (e.g., ``dataset'') to the user's intent. This query expansion simulates real-world search behavior, preventing a trivially weak baseline by ensuring the agent successfully targets data rather than general web content. The retrieved payload consists of URLs and text snippets synthesized from page content. 
    \item \textbf{Semantic Agent (Structured Corpus):} The agent queries the Google Dataset Search index via a metadata search tool. Representing the structured ecosystem, this curated corpus aggregates all pages with \texttt{schema.org/Dataset} markup. To maintain quality, we use a classification model to filter out roughly 80\% of pages with invalid or misused annotations \cite{dataset_or_not}. The remaining corpus contains 90 million dataset metadata records. Because the ecosystem natively restricts results to dataset entities, the query expansion is redundant. Therefore, the agent extracts the core keyword to query the index directly without the need for query expansion.
\end{itemize}

To eliminate confounding factors, both agents are governed by a set of controlled variables: (1) both utilize the same agentic framework (ADK) and underlying Large Language Model (Gemini 2.5 Pro); (2) prompts restrict agents to returning only datasets present in their tool-execution results, paired with a model temperature of 0 to maximize factual adherence and reproducibility; (3) agents are limited to returning a maximum of three highly relevant datasets per query to standardize evaluation volume; (4) findings must be output as a JSON array with dataset name, and source URL to facilitate automated evaluation; and (5) agents must output a deterministic fallback (``No relevant datasets found.'') if no datasets are discovered.



\begin{table}[htbp]
\centering
\caption{Agent Comparison}
\label{tab:ecosystem_comparison}
\begin{tabular}{@{} p{3.5cm} p{4.2cm} p{4.2cm} @{}}
\toprule
\textbf{Component} & \textbf{Baseline Agent\xspace} & \textbf{Semantic Agent} \\
\midrule
\textbf{Agent / Model} & ADK Agent / Gemini 2.5 Pro & ADK Agent / Gemini 2.5 Pro \\
\addlinespace
\textbf{Target Index} & Google Search Index & Google Dataset Search Index \\
\addlinespace
\textbf{Inclusion Filter} & General web crawlability & Semantic metadata markup \\
\addlinespace
\textbf{Corpus Scale} & Billions of web pages & $\sim$90 Million dataset metadata \\
\addlinespace
\textbf{Retrieved Payload} & URL and synthesized text from web pages & URL and semantic metadata \\
\bottomrule
\end{tabular}
\end{table}

\subsection{Real-World Indices vs. Curated Corpora}

We deliberately chose live, web-scale indices (Google Search and Google Dataset Search) to evaluate agentic retrieval under authentic conditions. While we adopt a query set from an established benchmark—specifically NTCIR-16—to ensure a reproducible evaluation of genuine user intent, we explicitly bypass the benchmark's underlying static document collections. Rather than operating in artificially sanitized environments—such as the NTCIR-16 data collections themselves, fixed web crawls (e.g., ClueWeb22~\cite{clueweb22}), or centralized, well-formatted dataset repositories (e.g., Zenodo, Hugging Face)---our approach forces agents to navigate the messy, decentralized open web. Deploying these queries against billions of live documents proves the system's scalable, ``in-use'' viability. Furthermore, live indices capture newly published data and real-world \schema adoption, avoiding the staleness of static corpora.

\section{Evaluation Methodology}
\label{sec:evaluation_method}

To compare the two agents, we use a public benchmark for dataset retrieval, develop evaluation metrics that explicitly address the FAIRness of the datasets, and design the autoraters to ensure scalable, low-variance evaluations.

\subsection{Benchmark Dataset}

We evaluate our systems on the NTCIR-16 Data Search 2 dataset~\cite{tao2022overview}, an established ad-hoc dataset retrieval benchmark spanning multiple domains like economics, demographics, and public health. Specifically, we use queries from its Information Retrieval (IR) subtask.

From this subtask, we isolated the English keyword-based questions (N=58) from the IR subtask to simulate the authentic data discovery intent. Because these queries were crowdsourced from human workers tasked with expressing real-world information needs derived from actual web pages, they simulate authentic data discovery intent (e.g., ``air quality baltimore''). Unlike synthetic or verbose prompts, these human queries are frequently underspecified and context-dependent. Utilizing this query set forces the agents to navigate the genuine complexities of matching user intent to dataset metadata, providing a realistic assessment of how these systems will perform in production environments.


\subsection{Evaluation Metrics}

We evaluate the agents for autonomous workflows across three FAIR-inspired dimensions: discovering relevant web pages (Findable), ensuring programmatic access (Accessible), and verifying the computational utility based on page types (Interoperable/Reusable).



\begin{itemize}
    \item \textbf{Relevance (Findable):} Evaluates semantic alignment between the NTCIR-16 query and dataset scope, assigning a score from -1 to 2 (summarized in Table \ref{tab:relevance_rubric}). We deliberately score ``Exploratory Matches'' as Highly Relevant. In real-world data discovery, users frequently issue broad queries that lack explicit geographic, temporal, or demographic constraints. Returning a highly specific dataset in response to a broad, underspecified query proves the agent successfully navigated the correct semantic domain without violating any explicit user constraints. 
    \item \textbf{Data Accessibility (Accessible):} Measures data accessibility via a multi-level rubric (Table \ref{tab:accessibility_rubric}) to assess complexity of data extraction and the presence of technical barriers.
    \item \textbf{Dataset Page Type (Interoperable \& Reusable):} Measures computational utility by categorizing the dataset pages into a page type rubric (summarized in Table \ref{tab:dataset_type_rubric}).
\end{itemize}

\begin{table}[h!]
    \centering
    \caption{LLM Autorater Rubric for Scoring Query-to-Dataset Relevance.}
    \label{tab:relevance_rubric}
    \renewcommand{\arraystretch}{1.2}
    \begin{tabular}{c p{0.25\textwidth} p{0.55\textwidth}}
        \toprule
        \textbf{Score} & \textbf{Classification} & \textbf{Description} \\
        \midrule
        2 & Highly Relevant & Contains requested data (exact match or superset), or provides concrete data for broad, unconstrained exploratory queries. \\
        1 & Partially Relevant & On-topic but incomplete or mismatched (e.g., wrong year/location, or a narrow subset of an explicitly constrained query). \\
        0 & Irrelevant & Entirely off-topic or unrelated to the query. \\
        -1 & Not Applicable & The page is unreachable. The relevance cannot be evaluated. \\

        \bottomrule
    \end{tabular}
\end{table}

\begin{table}[h!]
    \centering
    \caption{LLM Autorater Rubric for Scoring Data Accessibility Levels.}
    \label{tab:accessibility_rubric}
    \renewcommand{\arraystretch}{1.2}
    \begin{tabular}{l p{0.25\textwidth} p{0.6\textwidth}}
        \toprule
        \textbf{Level} & \textbf{Classification} & \textbf{Description} \\
        \midrule
        6 & Machine-Readable & Contains direct links or export options to structured data files (e.g., .rdf, .ttl, .csv, .json) or API endpoints. \\
        5 & Structured & Data in structured formats (e.g., tables) is presented in a page requiring parsing but not NLP. \\
        4 & Unstructured & Data points (e.g., statistics, coordinates) are embedded in prose, requiring NLP for extraction. \\
        3 & Presentation-Bound & Data is confined to static images, non-machine-readable documents, or interactive visual interfaces \\
        2 & Non-Data & Page is reachable but lacks data. \\
        1 & Unreachable & Page is unreachable (e.g., 403/404 errors). \\
        \bottomrule
    \end{tabular}
\end{table}

\begin{table}[h!]
    \centering
    \caption{LLM Autorater Classification Rubric for Retrieved Dataset Page Types.}
    \label{tab:dataset_type_rubric}
    \renewcommand{\arraystretch}{1.2}
    \begin{tabular}{p{0.30\textwidth} p{0.60\textwidth}}
        \toprule
        \textbf{Classification} & \textbf{Description} \\
        \midrule
        DATA\_REGISTRY & Dataset landing pages with interoperable metadata records (e.g., DOIs, data dictionaries, provenance). \\
        RAW\_DATA & Machine-readable files (e.g., .rdf, .csv) or APIs. \\
        DATA\_EXPLORER & Interactive interfaces (e.g., dashboards, dynamic maps) or standalone charts without metadata. \\
        DATA\_NARRATIVE & Prose-heavy content using data to support storytelling (e.g., papers, news articles, reports). \\
        DISCOVERY\_PORTAL & Gateways or search catalogs for querying across multiple distinct datasets. \\
        NO\_DATA & General web content lacking a specific dataset focus (e.g., organizational homepages, marketing). \\
        UNREACHABLE & Page is unreachable (e.g., 403/404 errors). \\
        \bottomrule
    \end{tabular}
\end{table}

\subsection{Autorater Execution}
\label{sec:autorater-execution}

\subsubsection{Autorater Evaluation Framework}
We use Gemini 2.5 Pro as an LLM-judge to evaluate our three metrics.\footnote{prompts: \url{https://doi.org/10.6084/m9.figshare.32141311}} Rather than relying on agent-generated snippets, the autoraters evaluate live content extracted directly from the agent-provided URLs and serialized into Markdown. To ensure reproducibility and prevent evaluation drift from link rot, we freeze this extracted content at query time and provide it to the autoraters alongside the dataset payload from agent response.

\paragraph{Chain-of-Thought and Evidence Extraction.} To prevent evaluation hallucinations, autoraters follow a strict Chain-of-Thought (CoT) \cite{wei2022chain} protocol. Before outputting a final rating, they must extract explicit evidence (e.g., quotes) from the serialized snapshot and map it to the rubric. This ensures evaluations are grounded in the scraped content rather than the model's parametric memory.

\paragraph{Relevance Prompt Design.} Our relevance evaluation adapts the UMBRELA prompt~\cite{upadhyay2024umbrela}, leveraging its few-shot examples and structured reasoning to distinguish between superficial mentions and primary data sources. To align with the NTCIR-16 Data Search benchmark, we converted UMBRELA's original 1–3 scale to a 0–2 ranking. This maps our outputs directly to established human-labeled ground truths while preserving the framework's zero-shot reasoning strengths.

\subsubsection{Autorater Validation and Manual Review}
\paragraph{Autorater Validation.} 
To validate our autoraters, two authors independently annotated a diverse 11\% sample of query-retrieval pairs to establish a ``gold set.'' Prior to analysis, we excluded non-ordinal 'Unreachable' categories. We then mapped the autorater's classifications to progressive ordinal scales for relevance and accessibility and and derived the scale for computational utility directly from the dataset page type.
We evaluated agreement using Linear-Weighted Cohen's Kappa ($\kappa$) to penalize minor ordinal mismatches less severely than major ones. Human annotators achieved high reliability, with Kappa scores of 0.70 (Page Type), 0.93 (Accessibility), and 0.95 (Relevance), resolving discrepancies via consensus. Comparing the autorater to this gold set yielded strong Kappa scores of 0.73, 0.78, and 0.74, respectively, demonstrating close alignment with human judgment.

\paragraph{Manual Evaluation for Edge Cases.} \label{sec:manuale-eval} 
Approximately 31\% of retrievals were routed to a human evaluation pipeline (29\% from \baseline and 33\% from \structured). As discussed in Section \ref{sec:limitations}, our internal scraper sometimes encounters complex layouts or security policies, classifying these scraping errors as ``Undetermined'' and flagging them for review. To avoid selection bias from discarding these edge cases, two authors independently annotated this subset. We evaluated inter-rater reliability using Linear-Weighted Cohen's Kappa, achieving scores of 0.72 (Dataset Type), 0.88 (Accessibility), and 0.70 (Relevance). Following consensus resolution, these human-annotated labels were integrated into the final dataset, ensuring a complete and statistically sound evaluation of both actual data availability and accessibility.


\section{Results}
We now present the experimental  outcomes of our comparative analysis across the 58 English keyword-based queries from the NTCIR-16 dataset. The \baseline answered 56 queries, retrieving 164 datasets. The \structured answered 40 queries, retrieving 112 datasets. 

\subsection{Relevance} 
\begin{figure}[h!]
    \centering
    \includegraphics[width=0.8\linewidth]{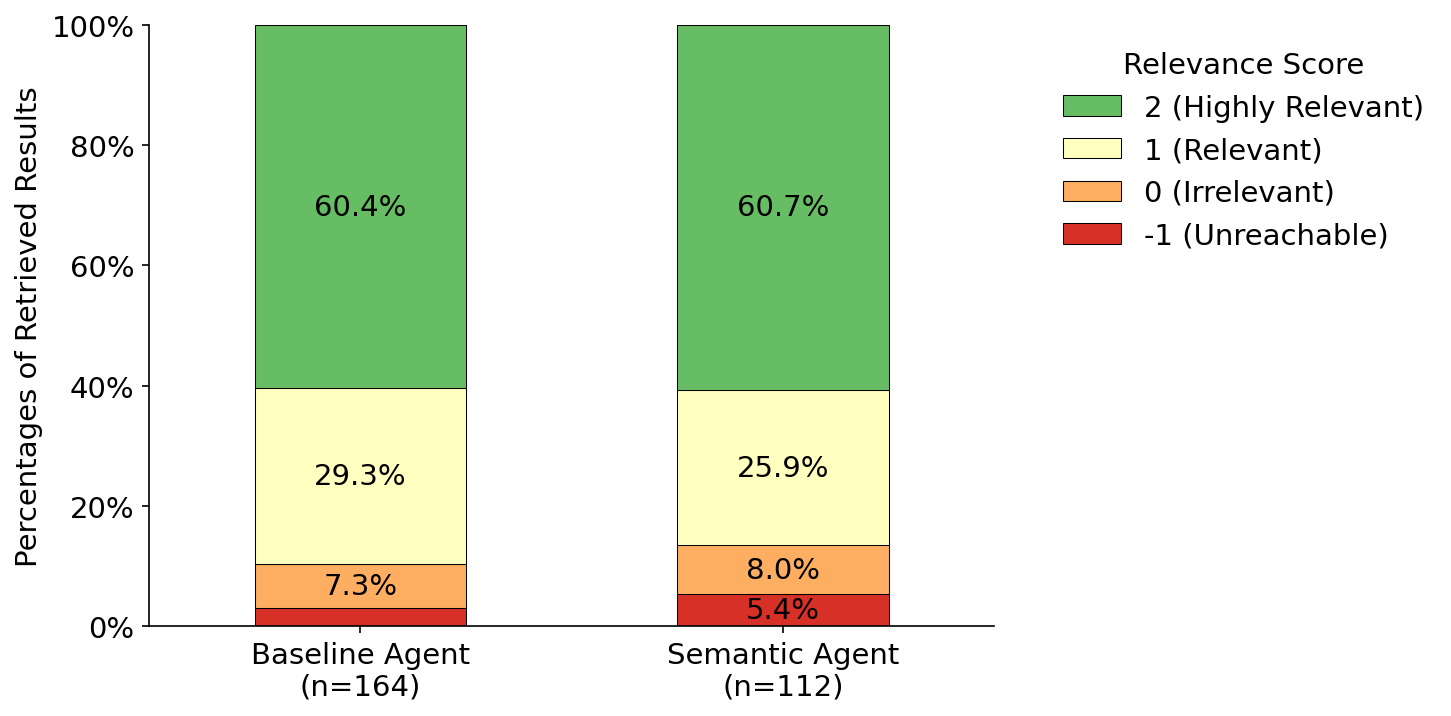}
    \caption{Comparison of the agent results by relevance scores.}
    \label{fig:relevance-results}
\end{figure}

Both systems exhibited similar performance in retrieving datasets (Figure~\ref{fig:relevance-results}) relevant to the questions. For the ``Highly Relevant'' (Score 2), the \baseline retrieved 99 datasets (60.4\%), while the \structured retrieved 68 datasets (60.7\%).

\subsection{Data Accessibility}

\begin{figure}[h!]
    \centering
    \includegraphics[width=0.8\linewidth]{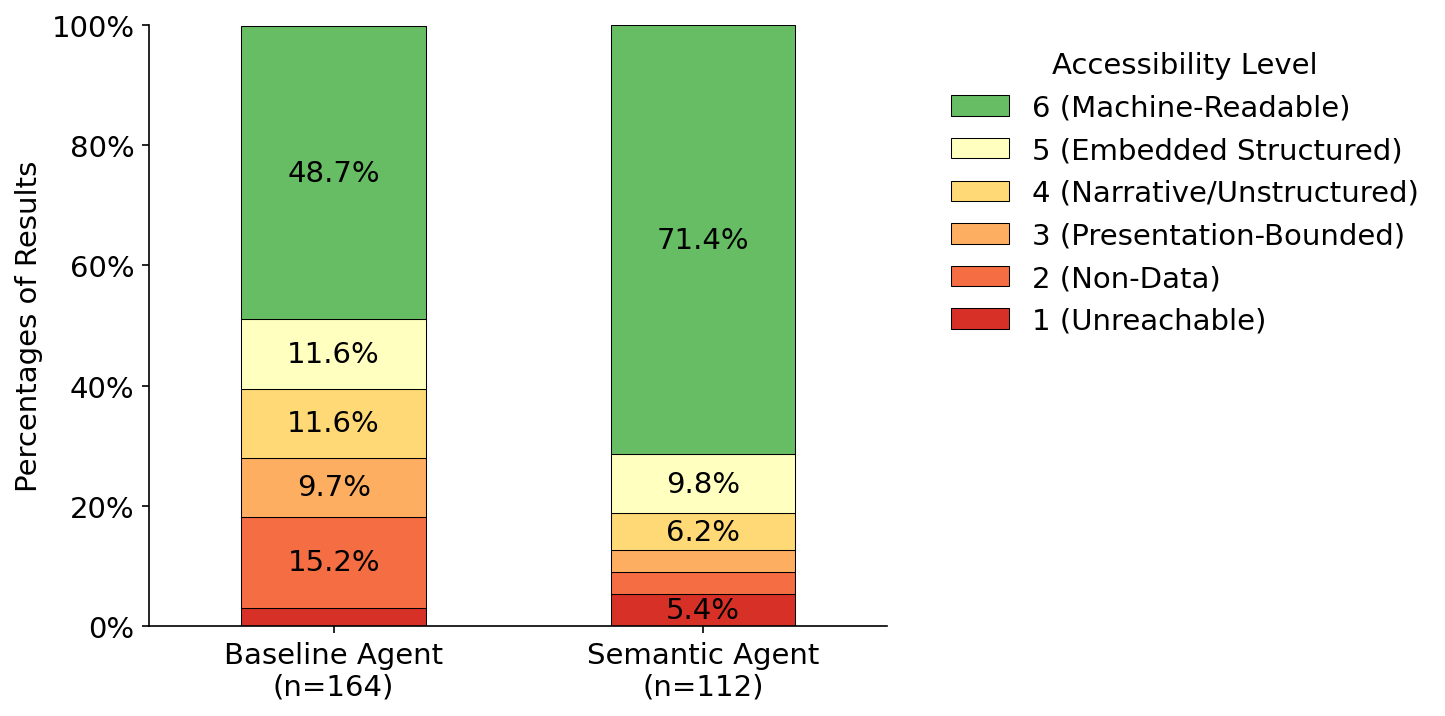}
    \caption{Comparison of the agent results by data accessibility levels}
    \label{fig:accessibility-results}
\end{figure}

The \structured demonstrated a significant accessibility advantage, returning pages with machine-readable data in 71.4\% of retrievals compared to 48.7\% for the \baseline. Accounting for the differences in total retrieval volumes, this represents a 46.6\% relative increase in precision (Figure~\ref{fig:accessibility-results}). By leveraging semantic metadata, the agent avoided non-computational roadblocks, achieving relative reductions of 46.6\% in Narrative/Unstructured Data (data embedded in narrative prose), 62.9\% in Presentation-Bounded Data (pages with only charts or interactive dashboards, without metadata), and 76.3\% in Non-Data (false positive pages lacking data entirely).



\subsection{Dataset Page Type}

\begin{figure}[h!]
    \centering
    \includegraphics[width=0.8\linewidth]{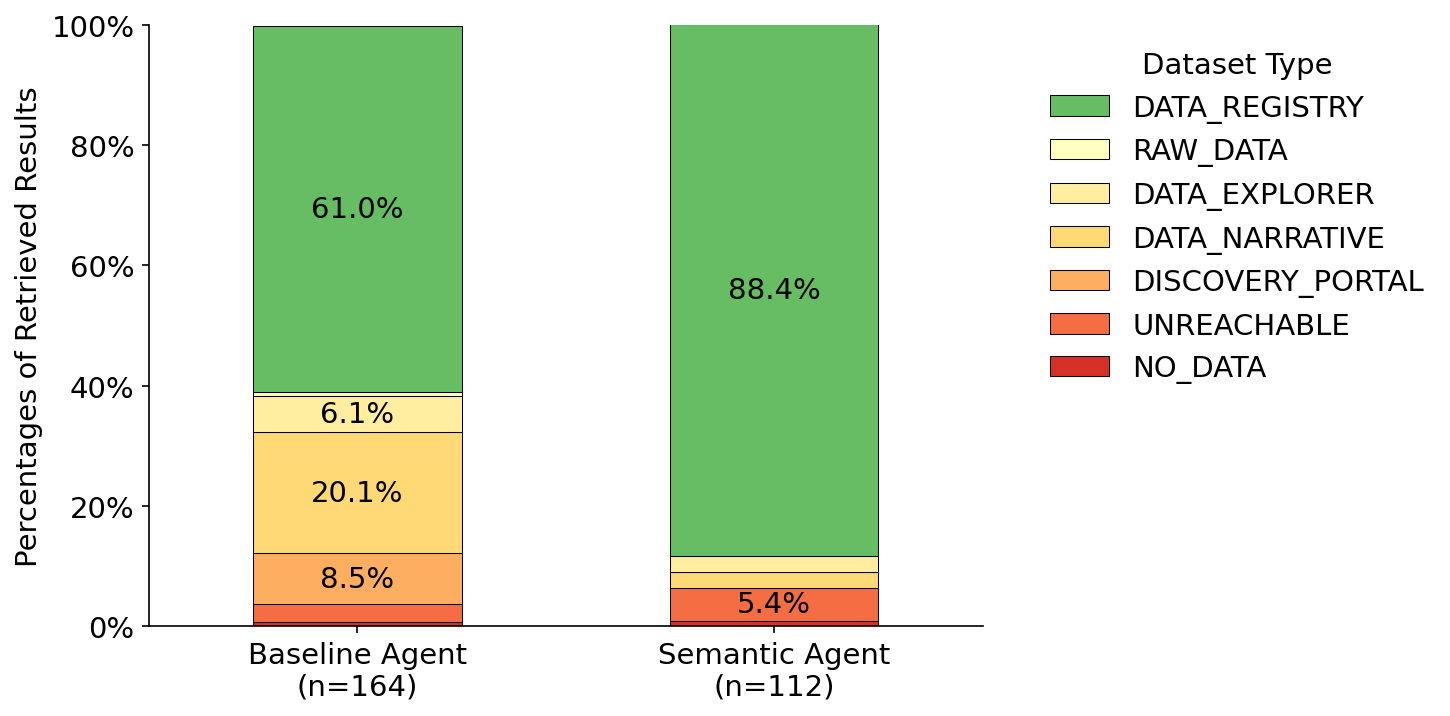}
    \caption{Comparison of the agent results by dataset page type}
    \label{fig:dataset-types-results}
\end{figure}

The distribution of retrieved dataset type pages varied between the two systems (Figure~\ref{fig:dataset-types-results}). The \structured's results consisted primarily of DATA\_REGISTRY entries. These accounted for 88.4\% of its total retrievals compared to 61.0\% for the \baseline, yielding a 44.9\% relative increase in precision. While the \baseline retrieved a wider variety of alternative page types, the \structured streamlined this distribution by achieving relative reductions of 86.6\% in DATA\_NARRATIVE pages (prose-heavy pages, such as news articles or academic papers discussing data), 55.7\% in DATA\_EXPLORER pages (data exploration pages without metadata), and 100\% in DISCOVERY\_PORTAL pages (search engines or gateways for multiple datasets).




\subsection{Agentic FAIRness}

To evaluate an agent’s capacity to identify machine-actionable registry entries that ensure metadata compliance and direct data accessibility. We define a dataset as fully ``FAIR-compliant'' if it achieves a perfect composite score across three criteria: a relevance score of 2 (Highly Relevant), Dataset Accessibility at Level 6 (Machine-Readable), and Dataset Page Type of DATA\_REGISTRY.

\subsubsection{Dataset-Level Precision}
To assess the overall precision of the retrieved data, we measured the dataset-level success rate---the proportion of FAIR-compliant dataset URLs out of the total number of URLs retrieved by each system across all queries. The \baseline achieved a 28.0\% precision rate (46 of 164 URLs). In contrast, the \structured achieved a precision of 46.4\% (52 of 112 URLs, $p < 0.01$). This represents a 65.7\% relative improvement, demonstrating that nearly half of the \structured's retrievals met the most strict FAIR criteria.

\subsubsection{Query-Level Result Density}
To evaluate the signal-to-noise ratio, we measured the concentration of FAIR-compliant datasets per answered query, capped at a maximum of 3 results per question. The \structured achieved a result density of 1.30 (52 FAIR-compliant dataset URLs / 40 answered queries), utilizing 43.3\% of its top-3 capacity. The \baseline managed a density of 0.82 (46 FAIR-compliant URLs / 56 answered queries), achieving a 27.4\% utilization rate. Due to the query size, this difference lacks statistical significance ($p > 0.05$).

\section{Discussion and Future Work}
Based on our findings, we revisit our core hypotheses, discuss practical strategies for autonomous agents navigating gated and constrained data, and propose a hybrid path balancing high-precision actionability with broad exploratory reach.

\subsection{Actionability, Utility, and Breadth}
We now return to our three key hypotheses in Section~\ref{sec:introduction}: data actionability, last-mile utility, and exploratory breadth. The \structured achieves superior precision by retrieving machine-actionable data for autonomous execution. Conversely, while the \baseline prioritizes exploratory breadth, it frequently fails at the ``last mile'' of data retrieval, where unstructured noise hinders autonomous action.

\subsubsection{Data Actionability} Our results validate the data-actionability hypothesis: the \structured demonstrated an advantage in navigating to these metadata-rich registries. More importantly, the \structured was more successful at identifying pages with machine-readable download links. 

\subsubsection{The ``Last Mile'' Utility}
The \baseline often retrieves prose-heavy web pages and navigational portals. Navigating and scraping these unstructured sources makes it harder for agents to isolate the actionable data. Specifically, web DOM syntactic noise causes severe token bloat, while dense prose degrades the context window, triggering retrieval failures \cite{li2026context}. Additionally, portals trap agents in redundant search loops instead of delivering data payloads directly\cite{aghzal2026agent}. Our results support this ``Last Mile'' utility hypothesis: while open-web agents can discover the relevant pages, they fail to locate actionable data.

\subsubsection{Exploratory Breadth}

The \baseline answered more questions than the \structured due to the limited adoption of \schema markup. This broader coverage is most pronounced in the web's ``long tail''— research domains and repositories where dataset pages often lack \schema annotations.

\subsubsection{Precision and Agentic Actionability}
In contrast to recall, the \structured method showed a clear precision advantage over the \baseline in retrieving FAIR-compliant datasets. For an autonomous agent, this advantage translates to a ``fail-fast'' mechanism that favors an empty state over a probabilistic guess, preventing downstream execution failures.



\subsection{Autonomy Strategy for Gated Data}
Real-world data discovery frequently encounters paywalls and authentication barriers. FAIR principles recognize these controls as standard research logistics rather than metadata failures---mandating that data remain ``as open as possible, as closed as necessary'' \cite{mons2017cloudy}. Rather than treating these barriers as absolute roadblocks, an autonomy strategy can employ Human-in-the-Loop (HITL) handoffs~\cite{pmlr-v162-humphreys22a}. An agent uses structured metadata to verify a dataset's utility, identifies the access blocker, and pauses to request human intervention for credentials or payment. Once the human clears the blocker, the agent resumes autonomously downloading and parsing the machine-readable payload.

\subsection{Scaling to Multi-Faceted Queries}
\label{sec:constrainted-based-search}


While our evaluation utilized keyword-based queries, it presents a compelling opportunity to explore complex, multi-faceted natural language queries. Evaluating the systems under metadata constraints exposes a contrast in their underlying mechanics. To satisfy specific constraints in an unstructured ecosystem, the \baseline must rely on probabilistic retrieval to identify candidate URLs, followed by high-friction scraping and parsing to verify attributes. Conversely, because structured corpora natively index explicit semantic properties (e.g., \texttt{license}, \texttt{dataType}), the \structured translates multi-faceted constraints directly into a native tool call with deterministic filter and reduces post-hoc verification overhead. As illustrated below, a multi-faceted natural language query can be converted into a metadata-constrained tool call:

\definecolor{SoftBlue}{HTML}{2B6CB0}
\definecolor{SoftPurple}{HTML}{805AD5}
\definecolor{SoftTeal}{HTML}{319795}
\definecolor{WarmOrange}{HTML}{DD6B20}

\begin{tcolorbox}[
    colback=gray!5,            
    colframe=gray!50,    
    top=5pt, 
    boxrule=0.5pt,             
    leftrule=3pt,              
    arc=0pt,                   
    outer arc=0pt,
    breakable
]
\textbf{Multi-faceted Query:}\\
``Find \textcolor{SoftBlue}{\textbf{tabular}} datasets on \textcolor{SoftPurple}{\textbf{household consumption}} data updated \textcolor{SoftTeal}{\textbf{within the last 3 months}} with \textcolor{WarmOrange}{\textbf{non-commercial}} licensing''

\vspace{0.5em}
\hrule
\vspace{0.5em}

\textbf{Metadata Tool call:}
\vspace{-0.3em}
\begin{flushleft}
\texttt{search\_datasets(}\\
\hspace*{2em}\texttt{keywords="\textcolor{SoftPurple}{\textbf{household consumption}}",}\\
\hspace*{2em}\texttt{last\_updated="\textcolor{SoftTeal}{\textbf{3 months}}",}\\
\hspace*{2em}\texttt{license="\textcolor{WarmOrange}{\textbf{noncommercial}}",}\\
\hspace*{2em}\texttt{dataType="\textcolor{SoftBlue}{\textbf{tabular}}")}
\end{flushleft}
\end{tcolorbox}


\subsection{Hybrid Path}

The choice to use semantic metadata should be driven by the agent's primary objective: breadth of knowledge versus reliability of action. Discovery-oriented tasks (e.g., exploratory research) require unstructured systems to achieve broad open-web recall, though this approach introduces higher noise that necessitates greater computational overhead or human intervention. Conversely, autonomous workflows (e.g., real-time analytics, code generation) require a semantic-metadata approach to guarantee machine-actionability, as false positives are far costlier than empty states.

To bridge the gap between reliability and breadth, we propose a hybrid architecture. Under this model, agents first query the high-precision semantic-metadata layer. If this initial search yields an empty state, the system falls back to the unstructured approach to cast a wider net.




%

\section{Limitations}
\label{sec:limitations}
 Our work has several limitations, specifically corpus coverage, ranking mechanisms, web scraping constraints, and queryset scale.
 
\paragraph{Coverage of the Structured Corpus.}
Our evaluation of the structured ecosystem is limited to the Google Dataset Search index (90 million records), which represents only a fraction of global data. Because inclusion strictly requires \texttt{schema.org/Dataset} or DCAT markup, our study excludes dataset pages using alternative markup vocabularies or lacking semantic annotation. Ultimately, this strict filtering ensures operational reliability but trades away  absolute recall and exploratory breadth of the open web.

\paragraph{Proprietary Ranking Mechanisms.}
Google Search and Google Dataset Search share the core infrastructure, but their internal ranking mechanisms remain ``black boxes.'' Rather than isolating this algorithmic delta, our evaluation measures end-to-end utility—assessing the empirical advantage of participating in the structured data ecosystem versus relying on the unstructured web.

\paragraph{Automated Scraping.}
\label{sec:automated_scraping}
Our internal scraper has limitations that prevent it from processing complex web layouts and penetrating enterprise-level security policies. As a result, the autoraters could not evaluate 31\% of the pages. We subsequently directed this subset to a human evaluation pipeline (\ref{sec:manuale-eval}) to mitigate bias, in order to prevent the scraping constraints from skewing evaluation results. Future work could mitigate this limitation by using advanced scraping tools.

\paragraph{Queryset Scale.}
Although our queryset yields statistically significant dataset-level metrics, its size may not fully capture the extreme diversity of the web's ``long tail.'' Furthermore, the query limits generalizability to the multi-faceted queries discussed in \ref{sec:constrainted-based-search}. Future work should employ larger, cross-disciplinary datasets to comprehensively evaluate agentic retrieval.

\section{Conclusion}
Autonomous workflows shift the primary focus of system utility from discovery to computational actionability. While unstructured retrieval supports broad exploratory tasks, structured ecosystems remain the indispensable foundation for reliable, execution-oriented autonomous workflows. In the agentic paradigm, the FAIR principles have transitioned from human-centric best practices into technical requirements. The  structured data ecosystem fuels the autonomous data-driven workflow, significantly increasing the probability that retrieved data assets are not just findable, but instantly machine-actionable.

\section{Use of Generative AI}

We used Gemini to generate initial drafts of some sections based on our discussion notes, as well as a copy-editing tool to improve grammar and overall readability throughout the manuscript. In addition, we used the model to critically review early drafts, incorporating its feedback to refine our arguments and restructure the narrative flow.




%
%
\bibliographystyle{splncs04}
\bibliography{references}
%






\end{document}

\end{document}